\def\+{{(+)}}  \def\-{ {(-)} }   \def\0{ {(0)} }
\def\1{ {(1)} }  \def\2{ {(2)} }

                       \def\th{\theta}

\documentstyle[11pt]{article}

\def\nn{\nonumber}
\def\be{\begin{equation}}             \def\ee{\end{equation}}
\def\ba{\begin{array}{rcl}}           \def\ea{\end{array}}
\def\beqa{\begin{eqnarray} }          \def\eeqa{\end{eqnarray} }
\def\beqalign{\begin{eqalign}}        \def\eeqalign{\end{eqalign}}

\def\bsubeq{\begin{subequations}}     \def\esubeq{\end{subequations}}
\def\bitem{\begin{itemize}}           \def\eitem{\end{itemize}}
\begin{document}
\title{\bf Real, $p$-Adic and Adelic  Noncommutative Scalar Solitons
\thanks{Talk presented at the Summer School in Modern Mathematical
Physics, Sokobanja and at the International Conference
FILOMAT2001, Ni\v s \ (Yugoslavia, August 2001). } }
\author{Branko DRAGOVICH$^{1,2}$\thanks{e-mail address: dragovich@phy.bg.ac.yu
}\ \ and Branislav SAZDOVI\'C$^1$\thanks{
       e-mail address: sazdovic@phy.bg.ac.yu}\\
      ${}^1$Institute of Physics, P.O.Box 57, 11001 Belgrade,
      Yugoslavia\\
      ${}^2$Steklov Mathematical Institute, Gubkin St. 8, 117966 Moscow,
Russia}
\date{}
\maketitle
\begin{abstract}
The actual interest in noncommutativity of space coordinates has
emerged recently in string theory as a consequence of the
properties of open string ending on D-branes.

This noncommutativity of coordinates $x$ induces the Moyal star
product $f \star g$ between (analytic) functions $f$ and $g$. To
investigate this subject in a systematic way, various
noncommutative field theory models are introduced in the last few
years. The basic equation in scalar field theory $\phi \star \phi
=\phi$  has an infinite number of solitonic solutions, unlike the
case with ordinary  multiplication.

The simplest solution of the above equation in the two spatial dimensions
is
$$
\phi_0(x)=2e^{-{x_1^2 +x_2^2 \over \theta}} \, .
$$
Introducing a new method, we reobtain this and other  solutions.
Also, we consider two $p$-adic   generalizations of the above
basic equation with the corresponding $p$-adic and adelic
solutions.
\end{abstract}

\section{Introduction} 

Let us consider single scalar field theory in $2+1$ dimensions
with coordinates $x^\mu$, $\mu=0,1,2 \quad (i=1,2)$. We suppose
noncommutativity in the spatial directions $[x^1, x^2]=i
\th^{12}=i\th$. The action  \cite{1} is \be  \label{1.1}  S=\int
d^3x \left[ {1\over 2} (\partial_0 \phi)^2 -{1\over 2} \partial_i
\phi
\partial^j \phi -V(\phi)_\star \right],
\ee where the potential \be \label{1.2} V(\phi)_\star ={m^2 \over
2} \phi^2 +\sum_n {a_n \over n} (\phi^n)_\star \, ,  \qquad
(\phi^n)_\star=\phi \star \phi \star
 \cdot\cdot\cdot\star \phi \, ,
\ee is defined in terms of star product \be  \label{1.3} (f\star
g)(x)=e^{{i \over 2}\theta^{ij}\partial_{a_i}
\partial_{b_j}} f(x+a) g(x+b)|_{a=b=0} \, ,
 \qquad (\partial_{a_i}= {\partial\over \partial a_i}) .
\ee Expanding the exponent and reordering the summation variables
we get \be  \label{1.4}  (f\star g)(x)= \sum_{k=0}^\infty
\sum_{q=0}^\infty \left({i\theta \over 2}\right)^{k+q} {(-1)^k
\over k! q!}
\partial_1^q \partial_2^k f(x) \partial_1^k \partial_2^q g(x) \, .
\ee
>From the property of the star product $\int d^2x
f\star g = \int d^2x f g$ it is clear that in the quadratic part
of the action the star product reduces to the usual one, so we
omit $\star$ in the first two terms in the action $S$.

We are interesting in the existence of {\bf solitons} in this
theory, i.e in  {\it time independent} solutions of the equation
of motion  when {\it energy} is finite. In the commutative limit
$\th=0$ the Derrick theorem states that in a pure scalar field
theory with $d\geq2$ ($d$ is a number of spatial dimensions) there
are no stable finite energy solutions (solitons). The proof of the
theorem is based on the rescaling of coordinates, but this
argument  fails in the presence of length scale $\sqrt{\th}$.

We will consider large noncommutative limit $\th \rightarrow
\infty$ in the energy \be E=\int d^2x [{1\over 2} (\partial_0
\phi)^2 + {1 \over 2} \partial_i \phi
\partial^j \phi +V(\phi)_\star] \,  ,   \ee which in the
static case reduces to the expression \be E=\int d^2x [ {1 \over
2}\partial_i \phi \partial^j \phi +V(\phi)_\star] \,.  \ee In
terms of the dimensionless coordinates $y={x \over \sqrt{\th}}$
the static energy becomes \be E=\int d^2y [ {1 \over 2} \partial_i
\phi \partial^j \phi +\th V(\phi)_\star]  \, ,  \ee where now
integration and partial derivatives are with respect to $y$. If
$\th \rightarrow \infty$, which in our case means $\th V \gg
\partial_i \phi \partial^j \phi$, the expression for the energy
reduces to the simple equation \be E=\int d^2y \th V(\phi)_\star
\, .       \ee The extremum of the energy corresponds to the
solution of the equation \be \label{1.9} {\partial V(\phi)_\star
\over
\partial \phi } =0 \,  .  \ee Note that in a commutative
case $(\th=0)$ , the above equation is a polynomial one, so the
solutions are only constants $\phi=\lambda_i \in \Lambda \, ,
\Lambda=\{ \lambda_1,...,\lambda_k\}$. Due to the presence of the
derivatives in the definition of the star product,  we expect here
some nontrivial solitonic type solutions.

Let us first consider the equation \be \label{1.10} F(\phi)_\star
=0 \, , \ee where $F(x)$ is an arbitrary analytic function and the
star means that fields in $F$ are multiplied using the star
product. If we are able to solve the equation \be \label{1.11}
\phi \star \phi = \phi \,  , \ee then we have  $\phi^n_\star=\phi$
and consequently $F(\lambda \phi)_\star =\sum_{n \geq 0}
{\lambda^n \over n!} F^{(n)} (0) \phi^n_\star =F(\lambda) \phi$.
So, $\lambda_i \phi$ is a solution of the equation (\ref{1.10}) if
$\lambda_i$ is a root of $F$. Similarly $\lambda_i \phi$ is a
solution of (\ref{1.9}) if $\lambda_i$ is an extremum of
$V(\lambda)$.

\section{Simple nontrivial solution} 

We are looking for a simple non-trivial solution of the basic
equation (\ref{1.11}). Let us check that the expression \be
\label{2.12} \phi_0(x)=2e^{-{x_1^2 +x_2^2 \over \theta}} \, ,  \ee
is a solution of this equation. We will call it a {\it simple
nontrivial solution}. Using the definition of the Hermite
polynomials \be \label{2.13} H_n (\alpha) = (-1)^n e^{\alpha^2}
\partial^n_\alpha  e^{-\alpha^2} \,, \qquad  (n=0,1,...) \,
 \ee we have \be \label{2.14}
\partial^n_\alpha  e^{-\alpha^2 \over \theta}=(-1)^n
\theta^{-{n \over 2}} e^{-\alpha^2 \over \theta} H_n\left({\alpha
\over \sqrt{\theta}}\right)\, ,  \ee and according to (1.4), the
star product can be written as \be \label{2.15} (\phi_0 \star
\phi_0)(x) =K\left(-i,{x_1 \over \sqrt{\theta}}, {x_2 \over
\sqrt{\theta}}\right) K\left(i,{x_1 \over \sqrt{\theta}}, {x_2
\over \sqrt{\theta}}\right) \phi_0^2(x) \, ,  \ee  where \be
\label{2.16} K(z,u,v)=\sum_{n=0}^\infty {z^n \over 2^n n!} H_n(u)
H_n(v) \, . \ee With the help of the representation of Hermite
polynomials \be \label{2.17} H_n(\alpha) ={ 2^n \over \sqrt{\pi}}
\int_{-\infty}^{\infty} dt e^{-t^2}  (\alpha+it)^n \, , \ee we
obtain \be \label{2.18} K(z,u,v)= { 1 \over \pi}
\int_{-\infty}^{\infty}dt e^{-t^2}
 \int_{-\infty}^{\infty} ds e^{-s^2+2z(u+it)(v+is)} \,   ,
\ee After integration over $s$, the coefficient in front of $\,
-t^2$ in the exponential turns into $1-z^2$, so the integration
over $t$ gives the following result \cite{2}  \be \label{2.19}
K(z,u,v)= {1 \over \sqrt{1-z^2}}e^{v^2 -{(v-zu)^2 \over 1-z^2}} \,
,  \ee under the condition $Re(1-z^2)>0$. Here $1-z^2 =2>0$ so
that \be \label{2.20} K(\mp i,u,v)= {1 \over \sqrt{2}}e^{{ u^2+v^2
\over 2}\mp iuv} \, ,
 \ee and consequently \be \label{2.21} K\left(-i,{x_1 \over \sqrt{\theta}},
{x_2 \over \sqrt{\theta}}\right) K\left(i,{x_1 \over
\sqrt{\theta}}, {x_2 \over \sqrt{\theta}}\right)= {1 \over
2}e^{x_1^2+x_2^2 \over \theta} ={1 \over \phi_0} \, .  \ee From
equations (\ref{2.15}) and (\ref{2.21}) it follows that $\phi_0$
is a solution of the (\ref{1.11}).

\section{ General solution} 

There exist infinitely many  solutions of the (\ref{1.11}). Let us
check that the expression \be \label{3.22} \phi_n(x)=(-1)^n
L_n\left(2{x_1^2+x_2^2 \over \theta}\right) \phi_0(x) \, ,  \ee
satisfies (\ref{1.11}). Here $L_n(t)$ are the Laguerre polynomials
defined as \be \label{3.23} L_n(\alpha)={1 \over n!} e^\alpha
\partial_\alpha^n(\alpha^n e^{-\alpha}) \, .
 \qquad  (n=0,1,...)
\ee

With the help of the following representation  \be \label{3.24}
L_n(\alpha)={1 \over n!}\partial_t^n\left({1 \over 1-t}e^{-{\alpha
t \over 1-t}}\right)\mid_{t=0} \, ,   \ee  instead of (\ref{3.22})
we can write \be \label{3.25} \phi_n(x)={(-1)^n \over
n!}\partial^n_t\left[{1 \over 1-t}\phi_0[\theta(t),x] \right] \mid
_{t=0} \, ,  \ee where $\phi_0[\theta(t),x]$ is the same
expression as $\phi_0(x)$ in   (\ref{2.12}), just substituting \be
\label{3.26} \theta(t)=\theta{1-t \over 1+t} \, ,  \ee instead of
$\theta$. In some sense, we expressed the $\phi_n$ in terms of
$\phi_0$, but with the $t$ dependent parameter $\theta$.

Now, we have \be \label{3.27} (\phi_n \star \phi_n)(x)= {1 \over
(n!)^2}
\partial_t^n \partial_s^n \left[ {1 \over
(1-t)(1-s)}\phi_0[\theta(t),x] \star \phi_0[\theta(s),x]
\right]\mid_{t=s=0} \, ,    \ee where the star product is defined
with respect to the parameter independent $\theta$. Similarly as
in the case with a simple nontrivial solution, using (\ref{1.4}),
(\ref{2.14}) and (\ref{2.16}) we find \beqa \label{3.28}
&&\phi_0[\theta(t),x] \star \phi_0[\theta(s),x]= K\left(-i
\sqrt{(1+t)(1+s) \over (1-t)(1-s)},
{x_1 \over \sqrt{\theta(s)}}, {x_2 \over \sqrt{\theta(t)}}\right) \nn \\
&&\times K \left(i \sqrt{(1+t)(1+s) \over (1-t)(1-s)}, {x_1 \over
\sqrt{\theta(t)}}, {x_2 \over \sqrt{\theta(s)}}\right) \,\,
\phi_0[\theta(t),x] \, \phi_0[\theta(s),x] .\,   \eeqa After some
calculations using (\ref{2.17}) we obtain \be  \label{3.29}
\phi_0[\theta(t),x] \star \phi_0[\theta(s),x]={(1-t)(1-s) \over
1+ts} \phi_0[\theta(-ts),x] \, .  \ee Note that because $z$ is
purely imaginary, the condition $Re(1-z^2)
>0$ is satisfied. Substituting this into (\ref{3.27}) we find
\be \label{3.30} (\phi_n \star \phi_n)(x)= {1 \over (n!)^2}
\partial_t^n
\partial_s^n \left[ {1 \over 1+ts} \phi_0[\theta(-ts),x]
\right]\mid_{t=s=0} \, ,   \ee and using the formulae \be
\label{3.31}
\partial_t^n \partial_s^n \varphi(-ts)\mid_{t=s=0}=
(-1)^n n!\partial_r^n \varphi(r)\mid_{r=0} \, ,   \ee we finally
obtain \be \label{3.32} (\phi_n \star \phi_n)(x)= (-1)^n {1 \over
n!}
\partial_r^n \left[ {1 \over 1-r} \phi_0[\theta(r),x]
\right]\mid_{r=0} \, ,    \ee which with help of (\ref{3.25})
yields expression (\ref{1.11}).

\section{$p$-Adic aspects}  

Since 1987, $p$-adic numbers and adeles have been successfully
applied in many topics of modern theoretical and mathematical
physics (for a review, see e.g. \cite{3,4,5}). In particular,
$p$-adic string theory \cite{3,4}, $p$-adic \cite{6} and adelic
\cite{7} quantum mechanics, as well as $p$-adic and adelic quantum
cosmology \cite{8,9}, have been formulated and investigated. It is
well known that combining quantum-mechanical and relativity
principles one concludes that there exists a spatial uncertainty
$\Delta x$ in the form  \be \label{4.33} \Delta x \geq \ell_0
=\sqrt{\frac{\hbar G}{c^3}} \sim 10^{-33} cm . \ee This
uncertainty relation (\ref{4.33}) may be taken as a reason to
consider simultaneously noncommutative and $p$-adic aspects of
spatial coordinates $x^i$ when approaching to the Planck length
$\ell_0$. Hence we are interesting here in $p$-adic analogs of the
above (real) noncommutative scalar solitons.

When we wish to consider basic properties of $p$-adic numbers it
is convenient to start with the field of rational numbers ${ Q} $,
since ${ Q}$ is the simplest field of numbers of characteristic
$0$ and it contains all results of  physical measurements. Any
non-zero rational number can be expanded  into two quite different
forms of infinite series. The usual one is to the base $10$, {\it
i.e.}
\begin{equation}
\label{dra2} \sum_{k=n}^{-\infty} a_k 10^k ,    \quad a_k
=0,\cdots, 9 ,
\end{equation}
and the other one is to the base $p$ ($p$ is a prime number) and
reads
\begin{equation}
\label{dra3} \sum_{k=m}^{+\infty} b_k p^k ,   \quad b_k =
0,\cdots, p-1 ,
\end{equation}
where $n$ and $m$ are some integers. These representations have
the usual repetition of digits, but  expansions are in the
mutually opposite directions.

The series (\ref{dra2}) and (\ref{dra3}) are convergent with
respect to the usual absolute value $\vert \cdot \vert_\infty $
and $p$-adic absolute value $\vert \cdot \vert_p$, respectively.
Allowing all possible combinations for digits, we obtain standard
representation of real numbers (\ref{dra2}) and $p$-adic numbers
(\ref{dra3}). ${ R}$ and ${ Q}_p$ exhaust all number fields which
contain ${ Q}$ as a dense subfield. They have many distinct
geometric and algebraic properties. Geometry of $p$-adic numbers
is the nonarchimedean one. For much more information on $p$-adic
numbers and $p$-adic analysis one can see,e.g. \cite{3,4,10,11}.

Practically, there are  two kinds of analysis on ${ Q}_p$ based on
two different mappings: ${ Q}_p\to { Q}_p$ and ${ Q}_p \to { C} $.
Both of them are used here. Elementary $p$-adic functions are
defined by the same series  as in the real case, but the region of
convergence is different. For instance, $\exp x =
\sum_{n=0}^\infty {{x^n}\over{n!}}$  converges if $\vert x\vert_p
< \vert 2\vert_p$. Derivatives of $p$-adic valued functions are
also defined as in the real case, but using $p$-adic valuation
instead of the absolute value.

Usual complex-valued $p$-adic functions are: {\it (i)} an additive
character $\chi_p (x) =\exp 2\pi i\{x\}_p$, where $\{x\}_p$ is the
fractional part of $x\in { Q}_p$, {\it (ii)} a multiplicative
character $\pi_s (x) = \vert x\vert_p^s$, where $s\in { C}$, and
{\it (iii)}  locally constant functions with compact support.
There is well defined Haar measure and integration.

In the previous sections, spatial coordinates  $x_1, x_2$ and the
noncommutativity parameter $\theta$ are real variables. In the
present section we are going to consider possible extensions of
the obtained results when these variables become $p$-adic valued.
One can introduce two types of the $p$-adic Moyal product. They
are $p$-adic differential and integral analogs of the usual Moyal
star product. While in the real case differential and integral
forms of the Moyal product are equivalent, we shall see that their
$p$-adic generalizations yield quite different expressions.

As a $p$-adic Moyal product in the differential form we take the
same expression (\ref{1.3}), where $f(x)$ and $g(x)$ are analytic
$p$-adic valued functions, and $\sqrt{ -1}$ is treated
$p$-adically. Note that \cite{3} $p$-adic exponential function
$\exp{x}$ is defined by the same infinite power series as in the
real case, but convergence is regarded with respect to the
$p$-adic norm and it is for $\exp{x}$ restricted to the domain
$|x|_p < |2|_p$. $p$-Adic derivatives are defined in the same
usual way but using $p$-adic norm instead of the absolute value.
Exponential derivative operator in (\ref{1.3}) acts formally in
the same way in real and $p$-adic cases. Now we are interested in
$p$-adic solution of the equation (\ref{1.11}) in the form
(\ref{2.12}), where $|x^i|_p < \sqrt{|2\theta|_p}, \ \ i=1,2$.
Expanding exponential function with derivatives one obtains \be
\label{4.36} (\phi_0 \star \phi_0) (x) = \sum_{n=0}^\infty A_n (x)
, \ee  where \be \label{4.37} A_n (x) = 4 \frac{i^n\theta^n}{2^n
n!} \left(\frac{\partial}{\partial x_1}\frac{\partial}{\partial
y_2} - \frac{\partial}{\partial x_2}\frac{\partial}{\partial y_1}
\right)^n \exp\left(-\frac{x_1^2 +x_2^2 +y_1^2 +y_2^2}{\theta}
\right) |_{y=x} . \ee

Performing some  calculations explicitly in (\ref{4.37}) one has
\be \label{4.38} A_n (x) = 4 \exp\left(-\frac{2(x_1^2
+x_2^2)}{\theta} \right) B_n(x),\ee  where $B_n (x)$ are: \beqa
\label{4.39} &&
 B_{2k+1}(x)=0,\ \ k=0,1,2,\cdots;\ \ \ B_0(x) = 1, \ \
  B_2 (x) = -1 + \frac{2}{\theta}(x_1^2 +x_2^2), \nn \\
&& B_4(x) = 1- \frac{4}{\theta}(x_1^2 +x_2^2) +
\frac{4}{2!\theta^2}(x_1^2 +x_2^2)^2, \nn \\ && B_6(x)= -1
+\frac{6}{\theta}(x_1^2 +x_2^2) -\frac{12}{2!\theta^2}(x_1^2
+x_2^2)^2 +\frac{8}{3!\theta^3}(x_1^2 +x_2^2)^3, \ \cdots \eeqa We
conclude that expansion (\ref{4.36}) is divergent since it
contains infinitely many divergent subseries like   \be
\label{4.40} 1-1+1-1+1-\cdots \ee Fortunately this subseries can
be easily made convergent using expression \be \label{4.41}
\sum_{k=0}^\infty (-1)^k q^{2k} = \frac{1}{1+ q^2} \ee and its
derivatives in the point $q = 1$. This kind of summation attaches
the sum $1/2$ to all divergent subseries and using this procedure
one gets that $ \phi_0 (x) =2\exp\left(-\frac{x_1^2
+x_2^2}{\theta} \right)$ is a solution of the equation $(\phi_0
\star \phi_0)(x) = \phi_0 (x)$, where $\star$ denotes the Moyal
product defined by (\ref{1.3}). It is worth noting that this
treatment of the equation $(\phi_0 \star \phi_0)(x) = \phi_0(x)$
does not depend on whether $x$ and $\theta$ are real or $p$-adic.
In other words, the above equation $(\phi_0 \star \phi_0)(x) =
\phi_0(x)$  and its solutions are number field invariant. Since
the sum of a divergent series depends on the way of summation it
follows that rearrangement in summation to get (\ref{1.4}) from
(\ref{1.3}) removed divergences in the real case of $(\phi_0 \star
\phi_0)(x) = \phi_0(x)$ in previous sections. However when $z= \pm
\sqrt{-1}$ the series (\ref{2.16}) is not convergent in the
$p$-adic case and such procedure is not suitable for $p$-adic
generalization.

In order to introduce the corresponding integral version of the
$p$-adic Moyal product let us recall that its integral form in the
real case is           \be \label{4.42} (f\star g)(x) = \int dk
dk' \exp\left( \frac{2\pi i}{h}(kx +k'x -\frac{\theta^{ij}}{2}k_i
k'_j) \right)\tilde{f}(k)\tilde{g}(k'), \ee where $\tilde{f}(k)$
and $\tilde{g}(k')$ are Fourier transforms $( \tilde{f}(k) = \int
\exp(-\frac{2\pi i}{h}kx)f(x)dx )$ of $f(x)$ and $g(x)$,
respectively. The equation (\ref{4.42}) is derived by the Weyl
quantization prescription of the functions on noncommutative
coordinates $\{\hat{x}^i  \}$, which satisfy canonical relation
\be \label{4.43}  [\hat{x}^i,\hat{x}^j]
=i\frac{h}{2\pi}\theta^{ij}. \ee The expression (\ref{4.42})  can
be rewritten  in the form  \be \label{4.44} (f\star g)(x) = \int
dk dk' \chi_\infty (-kx -k'x +\frac{\theta^{ij}}{2}k_i k'_j)
\tilde{f}(k)\tilde{g}(k'), \ee where $\chi_\infty (u) = \exp(-2\pi
i u)$ is the additive character in the real case, and it is also
taken $h=1$.

Defining the integral $p$-adic Moyal product as $p$-adic
generalization of (\ref{4.44})  we have  \be  \label{4.45} (f\star
g)(x) = \int dk dk' \chi_p (-kx -k'x +\frac{\theta^{ij}}{2}k_i
k'_j) \tilde{f}(k)\tilde{g}(k'), \ee where $p$-adic additive
character is $\chi_p (u)=\exp(2\pi i \{ u\}_p)$ and $\{ u \}_p$ is
rational part of $p$-adic number $u$. Note that  $f, g, (\chi_p) $
are real (complex) functions of $p$-adic variables and
$\tilde{f}(k) =\int \chi_p (kx)f(x) dx$. This $p$-adic integration
is well defined by the Haar additive measure.

The equation $(\varphi \star \varphi)(x) =\varphi (x)$ now reads
\be \label{4.46}  \int dk dk' \chi_p (-kx -k'x
+\frac{\theta^{ij}}{2}k_i k'_j)
\tilde{\varphi}(k)\tilde{\varphi}(k') = \varphi (x), \ee where for
simplicity we take $i,j = 1,2.$ According to the rules of
integration \cite{3} we find the following solution  \be
\label{4.47} \varphi_\nu (x) = \Omega (p^\nu|x_1|_p )\ \Omega
(p^{-\nu}|x_2|_p) \ee of (\ref{4.46}) with restriction
$\left|\frac{\theta^{ij}}{2}\right|_p \leq 1,$ where $\nu \in Z$
and \be \label{4.48} \Omega (p^\nu |x|_p) =  \left\{
\begin{array}{ll} 1, & |x|_p \leq p^{-\nu} ,   \\  0, & |x|_p >
p^{-\nu} .
\end{array} \right. \ee Note that the equation $ [\Omega (p^\nu
|x_1|_p)\ \Omega (p^{-\nu} |x_2|_p)]\star [ \Omega (p^\nu
|x_1|_p)\ \Omega (p^{-\nu} |x_2|_p)] =  \Omega (p^\nu |x_1|_p)\
\Omega (p^{-\nu} |x_2|_p)$ is also valid when $\star$ product is
replaced by the ordinary one. Since the Fourier transform of $
\Omega (p^\nu |x|_p)$ is  $ \Omega (p^{-\nu} |k|_p)$  we have   $
\tilde{\varphi}_\nu (k) = \Omega (p^{-\nu} |k_1|_p) \Omega (p^\nu
|k_2|_p) $. It means that when $\nu =0$ then function $ \Omega (
|x|_p)$ has its Fourier transform $ \Omega ( |k|_p)$, and it
resembles the same property of the Gaussian $\exp(-\pi x^2)$ in
the real case. In such sense function $ \Omega (|x|_p)$ is a
$p$-adic analogue of  $\exp(-\pi x^2)$. From the function
$\varphi_\nu (x) = \Omega (p^\nu|x_1|_p )\Omega (p^{-\nu}|x_2|_p)
$  and its Fourier transform it follows that $p$-adic solitonic
solution is in the region  $  |x_1| \leq p^{-\nu}, \quad |x_2|
\leq p^{\nu}, \quad |k_1| \leq p^{\nu}, \quad |k_2| \leq p^{-\nu}
$.

\section{Adelic aspects}

An adele  $x$  \cite{11} is an infinite sequence
\begin{equation}
\label{5.49}
   x = (x_\infty , x_2 , \cdots, x_p , \cdots) ,
\end{equation}
 where $x_\infty\in { R}$ and $x_p\in { Q}_p$ with the
restriction that for all but a finite set $S$ of primes $p$ we
have $x_p\in { Z}_p = \{x\in Q_p  : |x|_p \leq 1  \}$ . One can
use componentwise addition and multiplication. It is useful to
present the ring of adeles ${\cal A}$ in the following form:
\begin{equation}
\label{5.50}
  {\cal A} = \cup_S {\cal A}(S),
\end{equation}
\begin{equation}
\label{5.51} {\cal A}(S) = {R}\times \prod_{p\in S} { Q}_p \times
\prod_{p\not\in S} { Z}_p,
\end{equation}
where ${ Z}_p $ is the ring of $p$-adic integers. ${\cal A}$ is
also locally compact topological space with well-defined Haar
measure. There are mainly two kinds of analysis over ${\cal A}$,
which generalize those  over ${ R}$ and ${ Q}_p$.

Adelic approach gives possibility to treat real and all $p$-adic
aspects of a quantum system simultaneously and as essential parts
of a more complete description. Adelic quantum mechanics was
formulated \cite{7} and successfully applied to some simple and
solvable quantum models. Here we use adelic approach to
noncommutative scalar solitons.

According to the previous sections the equation $(\phi \star
\phi)(x) = \phi (x)$ can be considered as real as well as
$p$-adic. The $\star $ product may be realized in both
differential and integral form. Unlike the real case, $p$-adic
solutions of this equation depend on differential and integral
formulations. Hence we have two adelic versions of adelic aspects
of  $(\phi \star \phi)(x) = \phi (x)$, which are related to
differential and integral realizations of $\star$.

When $\star$ product has the differential (Moyal) form, we have
adelic valued solutions    \be \label{5.52} \phi (x) = (
\phi_\infty (x_\infty), \phi_2(x_2), \cdots, \phi_p (x_p), \cdots
), \ee where $\phi_\infty(x_\infty)$ and $\phi_p(x_p)$ are real
and $p$-adic valued functions of the form (\ref{3.22}),
respectively, and $|\phi_p(x_p)| \leq 1$ for all but a finite set
$S$ of prime numbers $p$. Possible region of $x_p$ is determined
by convergence of the exponential functions and it is $|x_p|_p
\leq \sqrt{|2\theta |_p}$. Parameter $\theta$ as a characteristic
of a physical system has to be rational, since a rational number
can be treated at the same time as real as $p$-adic.

If the $\star$ product has integral form then equation $(\varphi
\star \varphi)(x) = \varphi (x) $ has real valued solutions of
adelic variable $x$. These solutions have the following form: \be
 \label{5.53}   \varphi (x) = \varphi_\infty (x_\infty) \prod_{p\in S} [\Omega
(p^{\nu_p}|x^1_p|_p) \ \Omega (p^{-\nu_p}|x^2_p|_p)]
\prod_{p\notin S} [\Omega (|x^1_p|_p) \ \Omega (|x^2_p|_p)], \ee
where $S$ can be arbitrary finite set of primes $p$ and $\nu_p \in
Z$. In (\ref{5.53}) function $ \varphi_\infty (x_\infty)$ may be
any of the found real solutions (\ref{3.22}) of   the equation
$(\varphi \star \varphi)(x) = \varphi (x). $

{\bf Acknowledgements.\ } The work on this paper was supported in
part by the Serbian Ministry of Science, Technologies and
Development under contracts No 1426 and No 1486. The work of B.D.
was also partially supported by RFFI grant 02--01-01084  .

\end{document}